# Complex band structure and plasmon lattice Green's function of a periodic metal-nanoparticle chain

Kin Hung Fung, Ross Chin Hang Tang, and C. T. Chan

Department of Physics, The Hong Kong University of Science and Technology,

Hong Kong, China

When the surface plasmon resonance in a metal-nanoparticle chain is excited at one point, the response signal will generally decay down the chain due to absorption and radiation losses. The decay length is a key parameter in such plasmonic systems. By studying the plasmon lattice Green's function, we found that the decay length is generally governed by two exponential decay constants with phase factors corresponding to guided Bloch modes and one power-law decay with a phase factor corresponding to that of free space photons. The results show a high level of similarity between the absorptive and radiative decay channels. By analyzing the poles (and the corresponding residues) of the Green's function in a transformed complex reciprocal space, the dominant decay channel of the real-space Green's function is understood.

PACS number(s): 78.67.Bf, 73.22.Lp, 73.20.Mf, 78.70.-g

## I. INTRODUCTION

Band theoretic techniques are widely applied in describing the eigenmodes of periodic systems. With the help of the Bloch's theorem,[1] a band structure for propagating modes in a lossless periodic system is usually displayed as a plot of the eigenmode as a function of real frequency ($\omega$) versus real Bloch wave vector ($\mathbf{k}$) in the reduced-zone scheme.[2] Such a "real-$\mathbf{k}$" band-structure description has been generalized for including evanescent waves within bandgap by allowing the Bloch wave vectors to be complex quantities.[3,4] This generalized "complex-$\mathbf{k}$" description is sometimes called the complex-band-structure description.

There are several advantages of considering the complex band structure. For example, the magnitude of the imaginary part of the $\mathbf{k}$-vector gives the strength and the physical nature of the bandgaps. In addition, it can describe the decaying waves in lossy systems. It is generally believed that a non-zero imaginary part of $\mathbf{k}$ implies an exponential decay (or growth) of the wave with a decay constant given by $\ell = 1/\mathrm{Im}(k)$. Although the actual decay profile may depend on the incident field of a particular problem, such kind of exponential decay characteristics are indeed embedded in the Green's function.[5,6] For example, the frequency-domain Green's function for a homogeneous one-dimensional (1D) wave equation takes the form,[5] $G(\omega, x) \sim e^{i\mathrm{Re}(k)|x|} e^{-\mathrm{Im}(k)|x|}$, where $\omega$ is the excitation frequency, $k$ is the corresponding complex wave number, and $x$ is the spatial coordinate. Since the Green's function contains all the essential information on wave propagation, we can have a useful interpretation of the complex wave properties by studying the decay profile of the Green's function and its relation to the complex band structure.

Although there were some studies in the literature that is concerned with the Green's function for inhomogeneous lossy media, only a few of them focus on surface plasmons,[7,8] which have been shown to exhibit many interesting properties in recent years.[9] In plasmonic systems, which are intrinsically lossy due to absorption and radiation, the primary concern is the decay length of the plasmonic modes that can be excited in a certain point. In this paper, we study the Green's function for the plasmonic waves that propagate along a linear array of metal nanoparticles (MNPs).[10 11 12 13 14 15] We will first discuss the numerical results on the complex band structure (Sec. II) and the lattice Green's function (Sec. III) for such MNP array. Then, we will propose an *ansatz* which gives a closed-form expression for the lattice Green's function and we will show that the *ansatz* is indeed the correct solution by considering a generating function for the lattice Green's function (Sec. IV).

## II. COMPLEX BAND STRUCTURE

We consider a linear array of identical MNPs (with lattice constant $a$ and particle radius $r_0$). For $a \geq 3r_0$ and frequencies close to the Fröhlich frequency,[16] the electromagnetic modes of the MNPs can be accurately described by the coupled-dipole equations,[17]

$$\mathbf{p}_m = \alpha\left[\mathbf{E}_m^{ext} + \sum_{n\neq m}\vec{\mathbf{W}}(\mathbf{R}_m - \mathbf{R}_n)\mathbf{p}_n\right], \tag{1}$$

where $\mathbf{p}_m$ is the electric dipole moment of the $m$ th particle, $\alpha$ is the single-particle dynamic polarizability, and $\mathbf{E}_m^{ext}$ is the external electric field acting on the $m^{\text{th}}$ particle. The dynamic propagator, $\vec{\mathbf{W}}(\mathbf{r})$, for a dipole in a homogeneous host medium can be written as[17] $W_{uv}(\mathbf{r}) = k_h{}^3[A(k_h r)\delta_{uv} + B(k_h r)r_u r_v / r^2]$, where $k_h = \omega / c_h$, $c_h$ is the speed of light in the host medium, $A(x) = \left(x^{-1} + ix^{-2} - x^{-3}\right)e^{ix}$, $B(x) = \left(-x^{-1} - 3ix^{-2} + 3x^{-3}\right)e^{ix}$, and

$u, v = 1, 2, 3$ are component indices in Cartesian coordinates. Equation (1) can be written as

$$\sum_{n,v} M_{munv} p_{nv} = E_{mu}^{ext}, \tag{2}$$

where $p_{mu}$ and $E_{mu}^{ext}$ are the $u$ th components of the $\mathbf{p}_m$ and $\mathbf{E}_m^{ext}$, respectively. For a linear array of MNPs, Eq. (2) can be further split into decoupled equations in the form, $\sum_n M_{mn} p_n = E_m^{ext}$, for either the transverse modes (with $\mathbf{p}_m$ perpendicular to the chain axis) or longitudinal modes ($\mathbf{p}_m$ parallel to the chain axis). Since the transverse modes have richer properties in its band structure (such as negative slope[13] and strong coupling to free photon[7]), we will focus on the transverse mode.

The band structure can be calculated by setting $\mathbf{E}_m^{ext} = 0$ and finding pairs of $\omega$ and $k$ so that Eq. (2) has non-trivial Bloch solutions of the form $\mathbf{p}_m = \mathbf{p}_0 e^{ikma}$. More explicitly, we have to find the solutions of the equation,[14]

$$a^3 / \alpha(\omega) - \kappa(\omega, k) = 0, \tag{3}$$

where $\kappa(\omega, k) = k_h{}^2 a^2 \Sigma_1 + i k_h a \Sigma_2 - \Sigma_3$ for the transverse mode, $\Sigma_n = Li_n(e^{i(k_h - k)a}) + Li_n(e^{i(k_h + k)a})$, and $Li_n(x)$ is the polylogarithm function of order $n$. For lossless systems, $\omega$ and $k$ are real numbers. Since the system we consider here has both radiation loss and absorption loss, real values of $\omega$ and $k$ cannot satisfy Eq. (3). There are several ways to define the band structure of lossy systems but here we only focus on the complex band structure calculated by finding the complex $k$ that corresponds to each real $\omega$. A typical result calculated by numerical root searching in the complex-$k$ plane is shown in Fig. 1.

In the same figure, we compare the complex band structures for lossy ($\hbar\gamma$ = 0.02, results display as grey-colored lines) and lossless ($\hbar\gamma$ = 0, results display as blue-colored lines) metal with electric permittivity, $\varepsilon(\omega) = 1 - \omega_p^{\,2} / [\omega(\omega + i\gamma)]$, where $\gamma$ is the electron scattering rate. Chosen for an easy comparison with the results in the literature,[7 13] the plasma frequency for the Drude metal is $\hbar\omega_p$ = 6.18 eV and the geometrical parameters are chosen to be $a = 75$ nm and $r_0 = 25$ nm. Such parameters will reproduce the band structure in Ref. 13 outside the light cone (i.e., $k > \omega / c_h$) in the case of $\hbar\gamma$ = 0. The band structure has a negative slope for a range of $k$ close to the zone boundary. It should be noted that such a band of negative slope always exists when the particles are close enough. The slope of the band will turn to positive when the inter-particle distance becomes larger.

When there is no absorption, there is a critical frequency $\hbar\omega_c \equiv$ 3.272 eV below which there are two distinct guided mode solutions [$k_1$ and $k_2$, where $\mathrm{Re}(k_1) < \mathrm{Re}(k_2)$]. In this regime, the guided modes define a pass-band region because $\mathrm{Im}(k_1) = \mathrm{Im}(k_2) = 0$, indicating that the plasmon excitation is truly guided along the chain without radiation loss. As frequency increases to $\omega > \omega_c$, we have $\mathrm{Re}(k_1) = \mathrm{Re}(k_2)$ and $\mathrm{Im}(k_1) = -\mathrm{Im}(k_2)$. (This is due to the time reversal symmetry.) This region is a band-gap region in which the plasmon modes are leaky and can radiate to the surroundings. When we have $\hbar\gamma \neq 0$, all solutions have $\mathrm{Im}(k) \neq 0$. (The two solutions are distinct because the time reversal symmetry is broken by absorptive dissipation.) For $\omega < \omega_c$, the propagating plasmon experience a loss that is dominated by absorption. As frequency increases to $\omega > \omega_c$, both absorption and radiation losses can suppress the plasmon propagation. Details on the features of the band structure are available elsewhere[13] and we will not repeat the discussions. The aim of showing Fig. 1 is to show that the ratio between absorption and radiation losses can be adjusted by considering different frequencies and absorption rates. We can, therefore, find out the similarities and differences between the two kinds of losses by a suitable choice of parameters.

## III. LATTICE GREEN'S FUNCTION

We now consider the spatial decays of plasmonic excitations along the 1D array. The frequency-domain transverse lattice Green's function, $G(\omega, m, n) \equiv G(\omega, m-n)$, for the coupled dipoles satisfies

$$\sum_n M_{mn}(\omega) G(\omega, m-n) = \delta_{mn} . \tag{4}$$

The response dipoles for a given external driving field are thus given by $p_m = \sum_n G(\omega, m-n) E_n^{ext}$. $G(\omega, m)$ has the usual meaning of the response of the system when one particle is excited by a localized source ($E_n^{ext} = \delta_{n0} E_0^{ext}$) and can be written as a Fourier integral:[8]

$$G(\omega, m) = \frac{a}{2\pi} \int_{-\pi/a}^{\pi/a} \alpha_{eig}(\omega, k) e^{ikma} dk , \tag{5}$$

where $\alpha_{eig}(\omega, k) = [1/\alpha(\omega) - \kappa(\omega, k) a^{-3}]^{-1}$ is the Green's function in the reciprocal space.[18] It should be noted that $G(\omega, m) = G(\omega, -m)$ because of the mirror symmetry. While the complex band structure does not display the complete wave nature of the system (such as the exact response to the external field), the Green's function keeps all the information about the waves in the system. Therefore, all properties shown in the complex band structure can be fully explained by investigating $G(\omega,m)$ and we can see the actual meaning of the complex $k$ by comparing with $G(\omega,m)$.

There are at least two ways to evaluate $G(\omega,m)$. We can substitute the analytical formula of $\alpha_{eig}(\omega,k)$ into Eq. (5) and evaluate the integral. However, a closed-form solution to the integral is not available and numerical evaluation of the integral may run into stability problems.[8] Therefore, we choose to calculate, numerically, the inverse of $\mathbf{M}$ by truncating $\mathbf{M}$ to a matrix of finite dimension ( $N\times N$ ), which is the same as considering a finite chain of $N$ particles. In order to preserve the mirror symmetry, we consider $N$ as an odd number. Then, the Green's function is $G(\omega,m) =\sum_n \left(M^{-1}\right)_{mn} \delta_{n0} =\left(M^{-1}\right)_{m0}$, where $m= -N_0, -N_0+1, \ldots, N_0-1, N_0$ with $N_0=(N-1)/2$. Our method is similar to those in Refs. 7 and 8 except the single-site driving electric field is located at the center of the chain so that the mirror symmetry between the two ends of the chain is kept [see Fig. 2(a)]. In our calculations, we take $N = 3001$.

Here, we first briefly highlight some features of $G(\omega,m)$. When there is no absorption and radiation loss (i.e., $\gamma$ = 0 and $\omega<\omega_c$), $G(\omega,m)$ does not decay. As shown in Ref. 8, the results are intuitively obvious and thus the results are not repeated here. However, when the wave experience mainly absorption loss (i.e., $\gamma$ = 0.02 eV and $\omega<\omega_c$), the Green's function, $G(\omega,m)$, has two short-range exponential decays (which dominates for $m < 100$) and a long-range power-law decay ($m > 100$) [see Fig. 2(b)]. Due to the beating between modes of different wave numbers, we can observe oscillations at the crossovers between different decays. In the case of pure radiation loss (i.e., $\gamma$ = 0 and $\omega > \omega_c$), $G(\omega, m)$ has only one short-range exponential decay and a long-range power-law decay [see Fig. 2(c)]. In addition to the beating at the crossover points, there are strong oscillations for the whole exponential decay curve in Fig. 2(c). Such oscillations are associated with the time reversal symmetry and will be explained later in this paper. When radiation loss co-exists with absorption loss (i.e., $\hbar\gamma$ = 0.02 eV and $\omega > \omega_c$), the result [see Fig. 2(d)] shows no qualitative difference from that of Fig. 2(b) except that the exponential decay lengths are shortened. These interesting features (single/double exponential decays with power-law decay) have been largely ignored (though they indeed exist if we examine the details) in previous published works.[7] [8] In the following, we will understand such features by making an *ansatz* for the form of $G(\omega, m)$ and the accurate meaning of the complex Bloch wave vector will emerge accordingly.

## IV. ANSATZ AND JUSTIFICATIONS

We will use an *ansatz* of the lattice Green's function, $G(\omega, m)$ that takes the form,[19]

$$G_{ans}(\omega, m) \equiv c_1 z_1^{|m|} + c_2 z_2^{|m|} + f(m) z_0^{|m|}, \tag{6}$$

where $c_1$, $c_2$, $z_0$, $z_1$, $z_2$ are $m$-independent complex numbers satisfying $|z_2| \le |z_1| \le |z_0| = 1$ and $f$ is a bounded even function satisfying $\lim_{m\to\infty} m^p f(m) = 1$ for some real number $p$. The first two terms in Eq. (6) will give short-range exponential decays while the third term represents the long-range power-law decay. We will justify the *ansatz* in the following, by comparing $G(\omega,m)$ and $G_{ans}(\omega,m)$ and their generating functions in the complex number plane. By considering the generating functions, we will get extremely accurate values of the complex quantities (including $c_1$, $c_2$, $z_0$, $z_1$, and $z_2$) without using any function-fitting method.

To verify that $G(\omega,m)$ has to take the form as shown in Eq. (6), we consider the generating function of $G(\omega,m)$, defined as

$$g(\omega,z) \equiv \sum_{m=-\infty}^{\infty} G(\omega,m) z^m \ , \tag{7}$$

where $z$ is a complex number. By multiplying $z^m$ to Eq. (4) and summing the equation over the index $m$, one can show that the exact $g(\omega,z)$ reads

$$g(\omega,e^{ika}) = \alpha_{eig}(\omega,k) , \tag{8}$$

for any complex number $k$ by allowing analytic continuation (as in the usual definition of the polylogorithm functions). Thus, we have a closed-from expression for $g(\omega,z)$ and it has a one-to-one correspondence to $\alpha_{eig}(\omega,k)$, which is also the Green's function in the reciprocal space. To present a concrete picture, we plot $g(\omega,z)$ in the complex-$z$ plane (see Fig. 3) for various frequencies. From Fig. 3, we see that the exact generating function, $g(\omega,z)$, has 4 poles (at $z = z_1$, $z_1^{-1}$, $z_2$, and $z_2^{-1}$) and two branch cuts for each frequency. The poles are the solutions of Eq. (3) and the branch cuts, which show up as straight line segments, are defined by, $\mathrm{Arg}(z) = k_0 a$ for $0 < |z| < 1$ and $\mathrm{Arg}(z) = -k_0 a$ for $|z| > 1$. These branch cuts are the principle branch cuts of the polylogarithm functions. The advantage of considering $g(\omega,z)$ as a function of $z$ $(= e^{ika})$ instead of $\alpha_{eig}(\omega,k)$ as a function of $k$ is that all the information are compressed within a finite unit circle, $|z| \leq 1$, because of the mirror symmetry of the MNP chain [i.e., $g(\omega,z) = g(\omega,z^{-1})$]. In Figs. 3(a) and 3(b), we see that all four poles lie on the unit circle when there is no loss, i.e. $\omega < \omega_c$ and $\gamma = 0$. As we increase $\omega$ (keeping $\gamma = 0$), $z_1$ moves towards $z_2^{-1}$ while $z_2$ moves towards $z_1^{-1}$, and all four poles remain on the unit circle. After the poles meet at $\omega = \omega_c$ (not shown), they leave the unit circle separately [see Fig. 3(c) and 3(d)], keeping $|z_1| = |z_2| < 1$ (due to radiation loss). It should be noted that we have $z_1 = z_2^*$ (due to time reversal symmetry) when $\gamma = 0$. When $\gamma \neq 0$, the absorption breaks the time reversal symmetry, leading to $z_1 \neq z_2^*$ and smooth changes in the positions of the poles as frequency changes [see Figs. 3(e)-3(h)]. All these results can be directly mapped into the complex band structure and since each $z$ can be mapped into a wave number, $k$, the poles of the generating functions also give all the information required to produce the complex band structure in Fig. 1.

Let us now consider the corresponding generating function for our *ansatz* Green's function in Eq. (6), $g_{ans}(\omega,z) \equiv \sum_{m=-\infty}^{\infty} G_{ans}(\omega,m) z^m$ . Using Eq. (6), we can write

$$g_{ans}(\omega,z) = c_1 \left( \frac{z_1}{z - z_1} - \frac{z_1^{-1}}{z - z_1^{-1}} \right) + c_2 \left( \frac{z_2}{z - z_2} - \frac{z_2^{-1}}{z - z_2^{-1}} \right) + \sum_{m=-\infty}^{\infty} f(m) z_0^{|m|} z^m \,. \tag{9}$$

It should be noted that, in writing Eq. (9), we have again used the analytic continuation as in the polylogarithm functions. From Eq. (9), we see that the two short-range exponential decays in $G_{ans}(\omega,m)$ are associated with the four poles located at $z = z_1$, $z_1^{-1}$, $z_2$, and $z_2^{-1}$, which is consistent with $g(\omega,z)$ (as shown in Fig. 3). Here, we consider the exact values of the poles and the corresponding residues of $g(\omega, z)$. These quantities can be found accurately as long as we have rough estimates for the poles (for example, finding the poles from Fig. 3 by naked eye). With the estimated (and thus approximate) values of the poles, we can calculate a simple contour integral $R = \oint_C g(\omega, z) dz / 2\pi i$ numerically around a closed curve, $C$, that encloses the pole but not crossing the branch cuts. We thus obtain the four residues (denoted by $R_1$, $R_2$, $R_3$, and $R_4$) of $g(\omega, z)$ accordingly. It is expected that these four residues should be consistent with that of $g_{ans}(\omega, z)$. By equating the residues of $g(\omega, z)$ and $g_{ans}(\omega, z)$ (i.e. $R_1 = c_1 z_1$, $R_2 = -c_1 z_1^{-1}$, $R_3 = c_2 z_2$, and $R_4 = -c_2 z_2^{-1}$), we can solve for $c_1$, $c_2$, $z_1$, and $z_2$. The values of $z_1$ and $z_2$ obtained in this way are surprisingly accurate in the sense that they pin the poles of $g(\omega, z)$ with a relative error within $10^{-10}$ even when the relative error of our initial estimated values is about 0.1. This verifies that the form of the *ansatz* is essentially exact. To further make sure that the complex quantities, $c_1$, $c_2$, $z_1$, and $z_2$, can describe the initial exponential decays of $G(\omega, m)$ accurately, we also compare $c_1 z_1^m$ and $c_2 z_2^m$ with $G(\omega, m)$ [see Figs. 2(b), 2(c), and 2(d)]. [We have $c_1 = c_2^*$ and $z_1 = z_2^*$ in the case without absorption loss and thus $c_1 z_1^m$ and $c_2 z_2^m$ overlap in Fig. 2(c).] The results deduced from the *ansatz* show an exact agreement to $G(\omega, m)$.

We note that in addition to the exponential decay(s), there are oscillations in the magnitude of $G(\omega, m)$. We mentioned that those features near the crossover points are due to the beatings between different modes. Furthermore, there is an additional oscillatory behavior for the whole range of exponential decay in the case of no absorption because we have $c_1 z_1^m + c_2 z_2^m = 2\,\mathrm{Re}(c_1 z_1^m)$, which oscillates in its magnitude, when $\gamma = 0$. To verify these statements, we subtract both $c_1 z_1^m$ and $c_2 z_2^m$ from $G(\omega, m)$ and plot the remaining function, $G_{rem}(\omega,m) \equiv G(\omega,m) - c_1 z_1^m - c_2 z_2^m$, (dashed lines) in Fig. 2. We can see that the oscillatory features are immediately removed after the subtraction and the remaining function is smooth in its magnitude. This is another evidence to show that the *ansatz* gives the correct functional form of the Green's function.

Finally, we have to check whether $G_{rem}(\omega,m)$ is of the form, $f(m) z_0^m$, with $|z_0| = 1$. The verification of $z_0^m$ can be done by making a Fourier transform of the normalized function, $G_{rem}(\omega,m)/|G_{rem}(\omega,m)|$. We take a finite list of $G_{rem}(\omega,m)/|G_{rem}(\omega,m)|$ for $0 \le m \le 1000$ to calculate the Fourier transform numerically and find that the Fourier transform has only one single sharp peak located at $k \approx k_h$ (free-space wave number), which means we have $G_{rem}(\omega,m)/|G_{rem}(\omega,m)| \sim e^{ik_h ma}$. This is consistent with fixing $z_0 = e^{ik_h a}$ in our *ansatz*. Furthermore, if $G_{rem}(\omega,m) = f(m) z_0^m$, we will have $|G_{rem}(\omega,m)| = |f(m)|$. We can see in the insets of Figs. 2(b)-2(d) that $|G_{rem}(\omega,m)|$ is a smoothly decreasing function that behaves like $1/m^p$ for some constant $p$, manifesting as straight lines in the log-log plots (except for $m < 10$). Again, this verifies the form of the *ansatz*. The long-range power-law decay has a wave number very close to that of the free-space photons, suggesting that such portion of decaying wave is not confined within the chain and is less related to the guided modes. The power factor, $p$, is found to be close to 1.35, which is larger than the power factor for the Green's function of photons in free space.

## V. DISCUSSION AND CONCLUSION

In summary, we have shown semi-analytically that the (transverse) lattice Green's function for a periodic metal nanoparticle chain is of the form $G(\omega,m) = c_1 z_1^{|m|} + c_2 z_2^{|m|} + f(m)e^{ik_h|m|a}$ , which contains both exponential and power-law decays. This is independent of whether the decay is due to radiation or absorption loss. Our results show that a non-zero imaginary part of the Bloch wave number $k$ in the complex band structure only means an initial short-range exponential decay in the Green's function for such lossy system. There exists an additional power-law long-range decay that has no apparent relation to the values of $k$ of the guided mode. The existence of such power-law region suggests that extra care must be taken when one try to obtain the decay length by fitting the Green's function with exponential function. Our results can be helpful in understanding the wave propagation in lossy open systems. In addition, one interesting and subtle feature is that there are two exponential decay constants in the transverse mode, and it is true for frequencies that can excite guided modes or the leaky modes at the band gap above guided mode frequencies. The existence of two (instead of one) decay constants is due to the breaking of time reversal symmetry by absorptive dissipation. If there is no absorption, the two decay lengths become equal, as required by time reversal symmetry.

## ACKNOWLEDGMENTS

This work was supported by the Central Allocation Grant from the Hong Kong RGC through HKUST3/06C. Computation resources were supported by the Shun Hing Education and Charity Fund. We thank Dr. Dezhuan Han, Dr. Yun Lai, and Prof. Zhaoqing Zhang for comments and useful discussions.

201403(R) (2007); K. B. Crozier, E. Togan, E. Simsek, and T. Yang, Opt. Express **15**, 17482 (2007).

[12] M. Quinten, A. Leitner, J. R. Krenn, and F. R. Aussenegg, Opt. Lett. **23**, 1331 (1998); M. L. Brongersma, J. W. Hartman, and H. A. Atwater, Phys. Rev. B **62**, R16356 (2000); Stefan A. Maier, Mark L. Brongersma, and Harry A. Atwater, Appl. Phys. Lett. **78**, 16 (2001); S. Y. Park and D. Stroud, Phys. Rev. B **69**, 125418 (2004); D. S. Citrin, Nano Lett. **4**, 1561 (2004); R. A. Shore and A. D. Yaghjian, Electron. Lett. **41**, 578 (2005); J. J. Xiao, K. Yakubo, and K. W. Yu, Appl. Phys. Lett. **88**, 241111 (2006); A. F. Koenderink and A. Polman, Phys. Rev. B **74**, 033402 (2006); A. A. Govyadinov and V. A. Markel, Phys. Rev. B **78**, 035403 (2008); H. X. Zhang, Y. Gu, and Q. H. Gong, Chinese Phys. B **17**, 2567 (2008).

[13] C. R. Simovski, A. J. Viitanen, and S. A. Tretyakov, Phys. Rev. E **72**, 066606 (2005).

[14] A. Alù and N. Engheta, Phys. Rev. B **74**, 205436 (2006).

[15] K. H. Fung and C. T. Chan, Opt. Lett. **32**, 973 (2007).

[16] C. F. Bohren and D. R. Huffman, Absorption and Scattering of Light by Small Particles (Wiley, New York, 1983).

[17] V. A. Markel, J. Opt. Soc. Am. B **12**, 1783 (1995).

[18] The $\alpha_{eig}(\omega,k)$ defined here is the same as the eigenpolarizability defined in Ref. 15.

[19] The frequency dependence of every quantity (except $m$) on the right-hand side of Eq. (6) is not notated for the sake of a tidy expression.

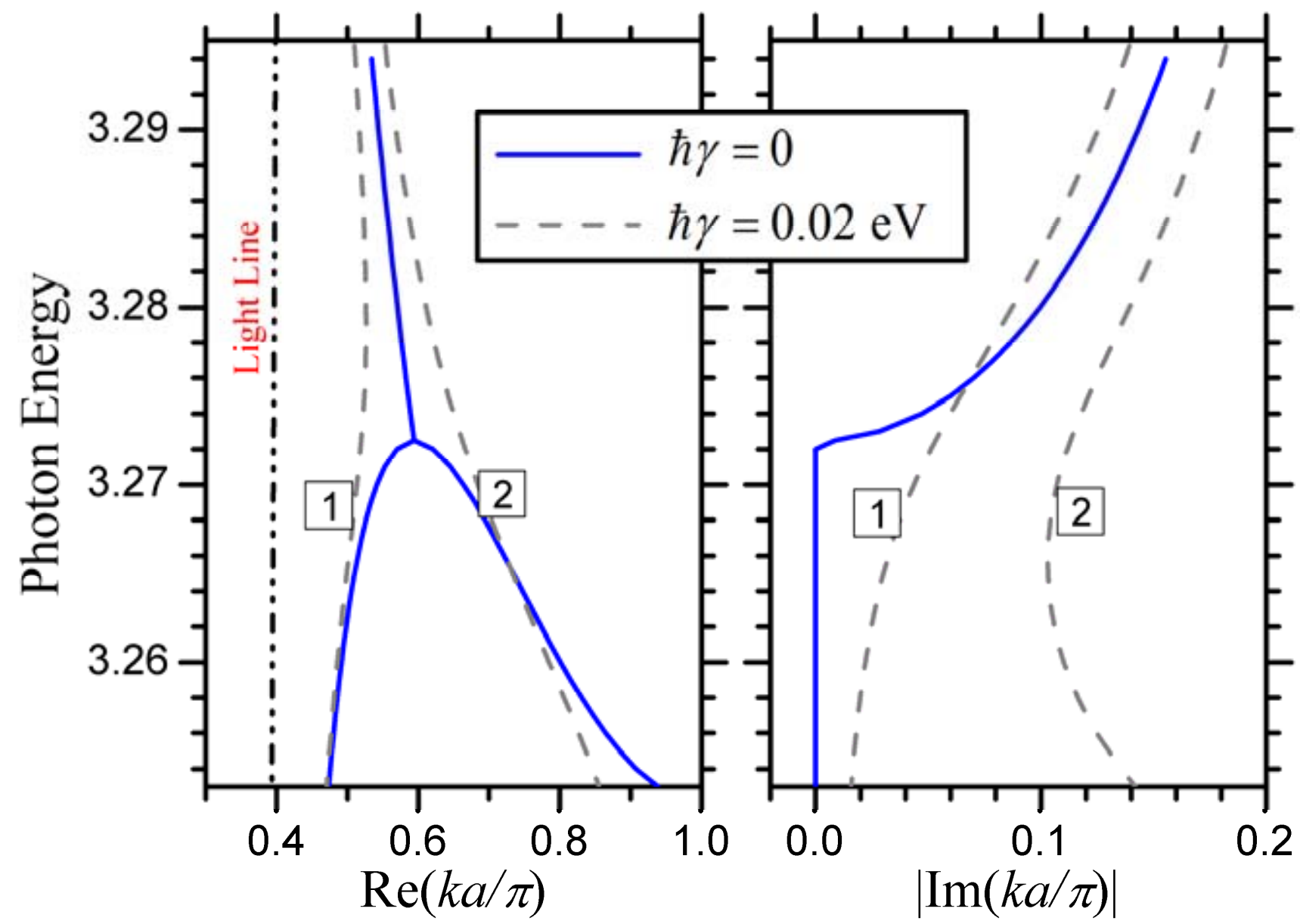

**FIG. 1: (Color online) Complex band structure for a 1D array of MNPs. Left (right) panel show the real (imaginary) part of the wave number. The numbers inside the square boxes indicate different non-degenerate modes when absorption is included.**

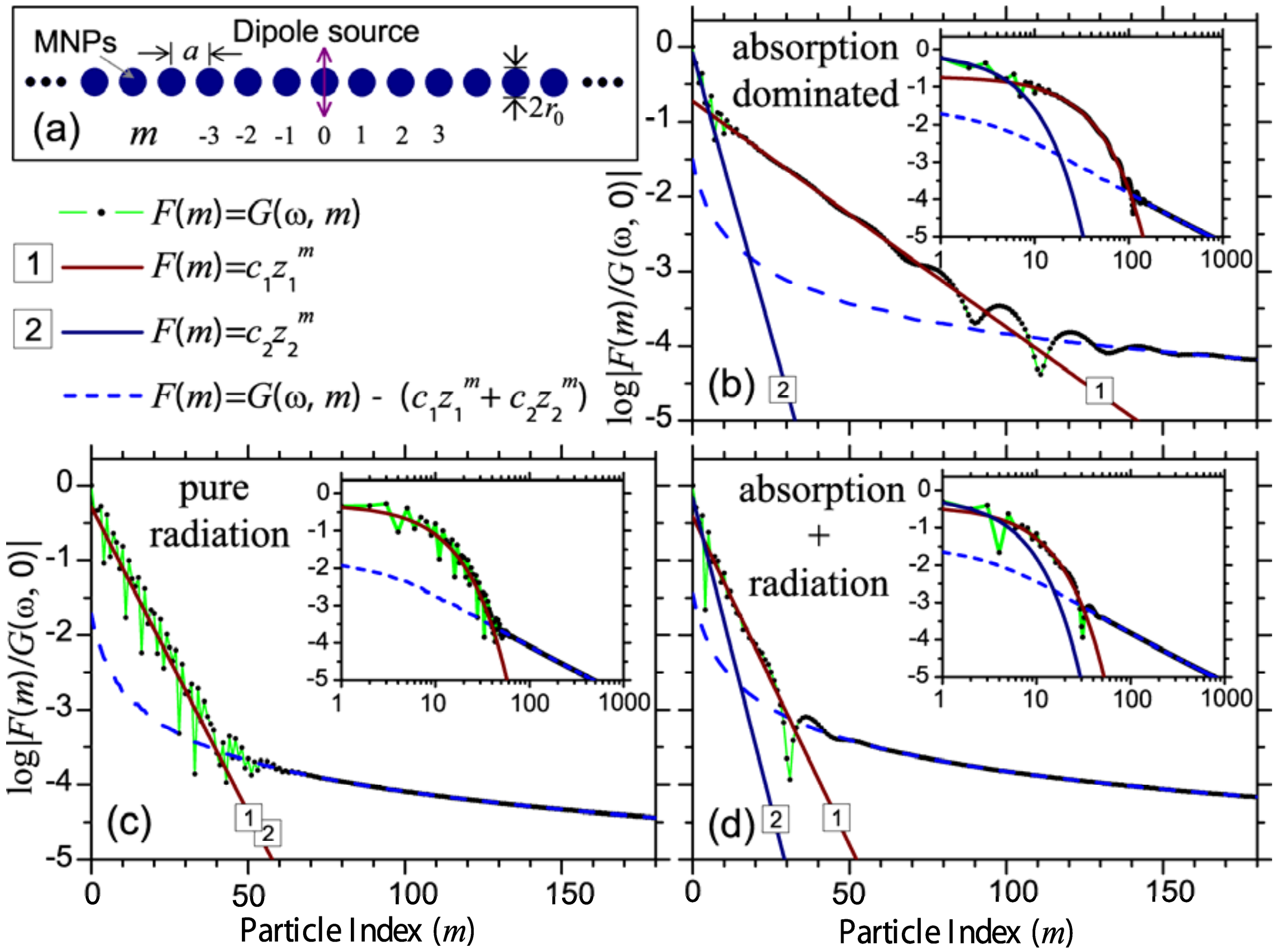

**FIG. 2: (Color online) Lattice Green's function $G(\omega,m)$. Panel (a) shows a schematic diagram with a dipole source located at the center of the MNP chain. Other panels show the values of $G(\omega,m)$ and the comparisons with the functions making up the *ansatz* for the case when energy loss is (b) dominated by absorption ($\hbar\omega$ = 3.260 eV and $\hbar\gamma$ = 0.02 eV), (c) purely due to radiation ($\hbar\omega$ = 3.275 eV and $\hbar\gamma$ = 0), and (d) contributed fairly by absorption and radiation ($\hbar\omega$ = 3.275 eV and $\hbar\gamma$ = 0.02 eV), respectively. The insets show the same graphs but in log-log scale, highlighting the long-range power law decay. Numbers with square boxes indicate the corresponding plasmon modes shown in Fig. 1.**

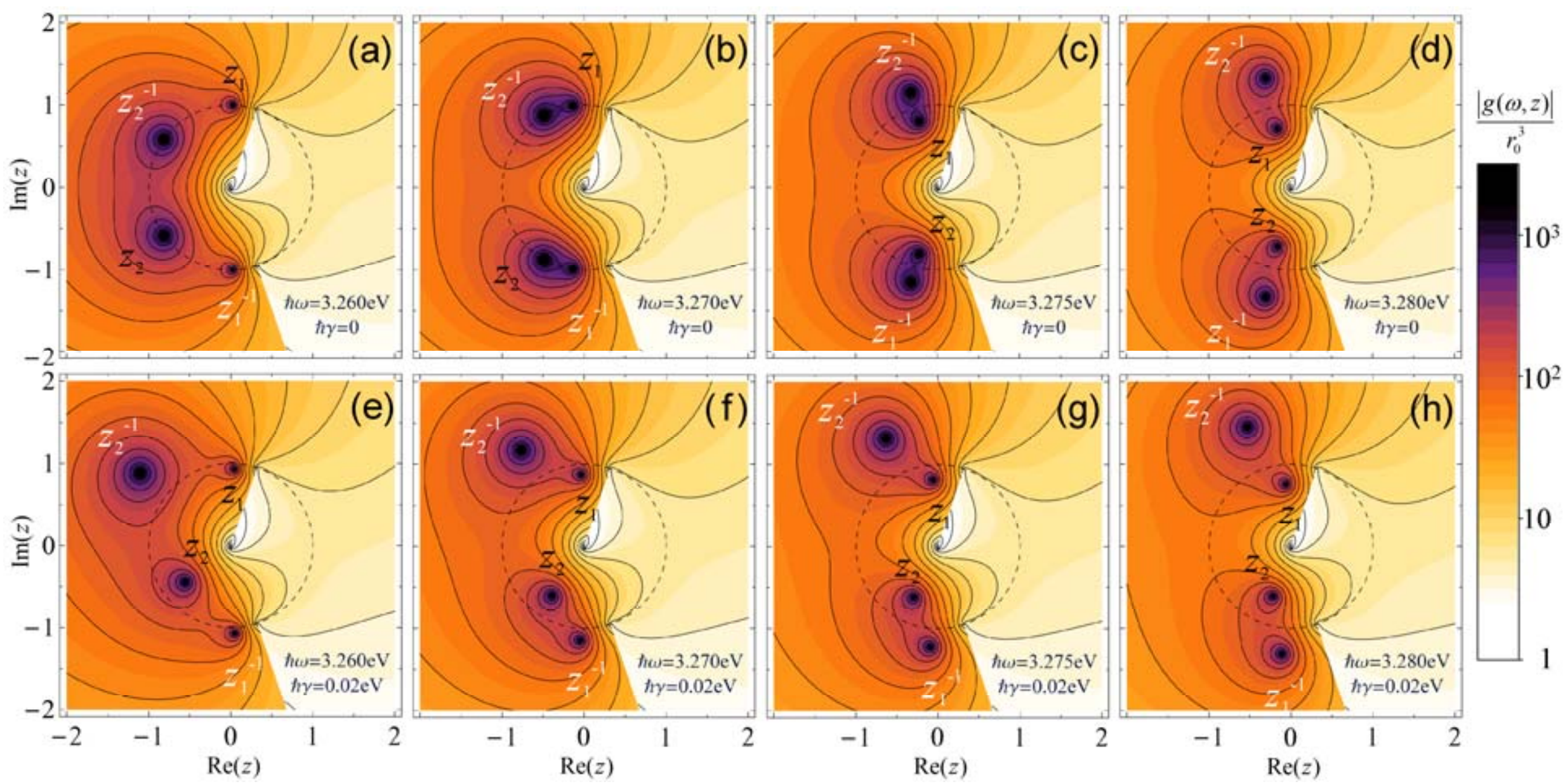

**FIG. 3: (Color online) Contour plot of $|g(\omega,z)|/r_0^3$ at different frequencies. The poles at $z=z_1$, $z_2$, $z_1^{-1}$, and $z_2^{-1}$ of the function are labeled. Upper (lower) panels show the results for the lossless (lossy) metal with $\hbar\gamma$ = 0 ($\hbar\gamma$ = 0.02 eV). (a) and (e) $\hbar\omega$ = 3.260 eV. (b) and (f) $\hbar\omega$ = 3.270 eV. (c) and (g) $\hbar\omega$ = 3.275 eV. (d) and (h) $\hbar\omega$ = 3.280 eV.**